# Interface Modification for Energy Levels Alignment and Charge Extraction in CsPbI3 Perovskite Solar Cells

Zafar Iqbal[1], Fengshuo Zu[2], Artem Musiienko[1], Emilio Gutierrez Partida[3], Hans Köbler[1], Thomas W. Gries[1], Gennaro V. Sannino[1,5], Laura Canil[1], Norbert Koch[1,2], Martin Stolterfoht[3], Dieter Neher[3], Michele Pavone[4], Ana Belen Muñoz-García[5], Antonio Abate[1,6,7*], and Qiong Wang[1*]

[1] Helmholtz-Zentrum Berlin für Materialien und Energie GmbH. Hahn-Meitner-Platz 1, 14109 Berlin, Germany.

[2]Institut für Physik & IRIS Adlershof, Humboldt-Universität zu Berlin, 12489 Berlin, Germany.

[3]Institute for Physics and Astronomy, University of Potsdam, Karl-Liebknecht-Straße 24−25,

14476 Potsdam-Golm, Germany.

[4]Department of Chemical Sciences, University of Naples Federico II, Comp. Univ. Monte S. Angelo, Via Cintia 26, 80126 Naples, Italy

[5]Department of Physics "Ettore Pancini", University of Naples Federico II, Comp. Univ. Monte S. Angelo, via Cintia 26, 80126 Naples, Italy

[6]Department of Chemistry Bielefeld University, Universitätsstraße 25, 33615 Bielefeld, Germany

[7]Department of Chemical Materials and Production Engineering, University of Naples Federico II, Piazzale Vincenzo Tecchio 80, 80125 Naples, Italy

## Abstract

In perovskite solar cells (PSCs) energy levels alignment and charge extraction at the interfaces are the essential factors directly affecting the device performance. In this work, we present a modified interface between all-inorganic CsPbI3 perovskite and its hole selective contact (Spiro-OMeTAD), realized by a dipole molecule trioctylphosphine oxide (TOPO), to align the energy levels. On a passivated perovskite film, by $n$-Octyl ammonium Iodide (OAI), we created an upward surface band-bending at the interface by TOPO treatment. This improved interface by the dipole molecule induces a better energy level alignment and enhances the charge extraction of holes from the perovskite layer to the hole transport material. Consequently, a $V_{oc}$ of 1.2 V and high-power conversion efficiency (PCE) of over 19% were achieved for inorganic CsPbI3 perovskite solar cells. Further, to demonstrate the effect of the TOPO dipole molecule, we present a layer-by-layer charge extraction study by transient surface photovoltage technique (trSPV) accomplished by charge transport simulation.

## Keywords:

Perovskite solar cells, surface band-bending, band energy alignment, charge selectivity, device stability, transient surface photovoltage technique (trSPV), Cyclic MPP tracking

## Introduction

Since Miyasaka *et al.* reported halide perovskites for the first time as a light absorber in solar cells in 2009,[1] the research community has improved the devices' PCE to 26%,[2] which makes perovskite solar cells (PSCs) potentially competitive with established technologies.[3-4] However, the stability of PSCs is one of Achille's heels to find the marketplace.[5-6] Inorganic perovskite $CsPbI_3$ has attracted attention due to its superior thermal stability compared to organic-inorganic lead perovskites.[7] In $CsPbI_3$-based PSCs, phase stability at the operating temperature and the crystallization processes are the main associated challenges with device fabrication.[8,9] Recently, organic precursors *e.g.* "$HPbI_3$" and DMAI-assisted film growth[10-11], and molten salt-based strategies have been adopted to address these challenges. To date, over 21% efficiency has been reported for $CsPbI_3$ with $V_{oc}$ of over 1.2V, [12] but inorganic perovskite solar cells have large open circuit voltage deficit as compared to organic-inorganic halide perovskite solar cells.[12,13] Currently, inorganic perovskite film post-treatment methods have been very effective to increase open-circuit voltage values.[14-16] However, the energy levels misalignment between the inorganic perovskite layer and the charge selective contacts requires interfacial engineering to resolve the open circuit voltage deficit.[17-19]

Inorganic perovskite has a flat band surface and introducing a surface band bending has been proven to be favorable for charge extraction.[20,21] L.Canil *et al*.,[22] systematically studied how the Fermi level of triple-cation perovskite can be tuned by the addition of dipole molecules on the surface. M. Nazeeruddin *et al* reported surface band bending on perovskite surface, they deposited perhydropoly(silazane) on top of triple-cation perovskite film [23] and found that efficient hole extraction is achieved due to surface band bending. Recently, Shuo Wang *et al.* introduced *n*-type surface band bending with propylamine hydrochloride molecule treatment for $CsPbI_3$ inverted devices.[21] This surface band bending impelled the better charge extraction and they have reported 20.17% state-of-the-art efficiency for inverted $CsPbI_3$ based solar cell. These reports suggest that surface band bending is an effective strategy to address the energy level alignment and charge extraction at the interfaces.

In this work, we used previously reported *n*-octyl ammonium iodide (OAI) passivation on $CsPbI_3$ to establish a control device. [24]. Passivating the surface defects of the perovskite has been reported systematically as one of the most effective strategies to enhance stability in PSC.[25-26] Long-chain alkylammonium halides were the most common molecules used for perovskite surface passivation. [27-28] However, we have revealed that OAI treatment leads to a downward band bending at the surface, which makes a more *n*-type perovskite film. For better alignment, we have introduced the dipole molecule trioctylphosphine oxide (TOPO) to induce a perovskite upward surface band bending without changing the concentration of surface defects on a well-passivated perovskite film. This improved interface sample is referred to as "with TOPO" (w/TOPO). TOPO treatment on the control film in *n-i-p* PSCs improved the charge selectivity six-fold, causing a decrease in energy offset and optimizing the energy levels alignment, significantly impacting the stability of state-of-the-art inorganic PSCs.

Non-radiative recombination and charge extraction simultaneously affect charge decay dynamics. Thus, the role of the TOPO molecule is challenging to differentiate by the photoluminescence methods due to quenching phenomena. As both charge extraction and passivation can contribute to quenching. Generally, the higher efficiency and stability were mainly attributed to the suppression of non-radiative recombination induced by surface defects.[29-30] Here we developed a method based on time-resolved Surface photovoltage [31] and charge transport simulation to resolve both carrier extraction and non-radiative recombinations. This method allows us to associate the improved stability of $CsPbI_3$ perovskite solar cells directly with the superior charge selectivity. Since we have separated the role of defect-passivation from upward surface band bending by adopting TOPO as a surface dipole on top of OAI for creating upward surface band bending. We also demonstrate the impact of the interfacial energy level alignment and the consequently improved charge extraction on the stability of PSCs.

**Results and Discussion**

With a bandgap of around 1.7 eV [32], $CsPbI_3$ has great potential in applying as a top cell in a tandem structure with silicon or a narrow bandgap perovskite film as the bottom cell.[33-34] $CsPbI_3$ perovskite thin films are prepared to adopt the method reported in one recent work from Soek's group.[24] Experimental details for perovskite film preparation and device fabrication can be found in the supporting information (SI). A potential influence of TOPO treatment on perovskite film was investigated by UV-Vis absorption, scanning electron microscopy, and Kelvin probe. The TOPO treatment barely changes absorption spectra, film thickness, and roughness as detailed in the SI. (**Fig. S1-4**).

We first conducted steady-state photoluminescence (PL) for neat films that are the control sample and treated with TOPO (w/TOPO) (**Fig.1a**). For TOPO-treated samples at varying concentrations of 5 mM, 10 mM, 15 mM and 20 mM, they exhibited similar photoluminescence quantum yield and thus a similar quasi-Fermi level splitting (**Fig. S5-I**). This indicates that TOPO treatment has barely any contribution to defect passivation. Time-resolved photoluminescence (TrPL) spectra, given in **Fig. 1b**, were fitted with the biexponential equation as discussed in SI **(Fig. S6)**. The fitted lifetime, $t_2$, referring to the non-radiative recombination, showed that all samples have a very similar value of around 3.5 μs. Thus, it further supports the steady-state PL data that TOPO treatment does not contribute noticeably to defect passivation.

X-ray photoelectron spectroscopy (XPS) results (**Fig. 1c, d**) revealed that the OAI treatment shifted the Pb 4f and I 3d spectra towards higher binding energy by 250 meV and 200 meV, respectively. This is in line with the previous report and is likely due to the stronger coordination of the Pb-I bond with the introduction of a long-chain cation.[11] The change of work function position within the band gap can be determined from the shifts of occupied electronic states (valence and core levels). Analysis of core levels provides direct information on the change of the chemical states, but the valence bands are overlapped with contributions from perovskite and top layers which render the analysis difficult. This is reflected as a less sharp valence onset in the valence spectra given in **Fig.2b**. As a result, we estimated the work function change within the band gap, from the core level electrons binding energy shift of I 3d (as it has lower kinetic energy of the photoelectrons as compared to that of Pb 4f, which probes the magnitude of surface band bending more precisely), that leads to a downward band bending at the surface by 200 meV.

With further TOPO treatment, Pb 4f and I 3d peaks are shifted to lower binding energy by 80 meV and 100 meV respectively. Thus, it indicates a lower downward band bending at the surface by 100 meV. We suggest here the core level binding energy shift in Pb 4f and I 3d comes from the electron-donating nature of TOPO [35]: the ligand may donate electrons to positively charged ions, such as $Pb^{2+}$ or iodine vacancies [36] on the perovskite surface, thereby leading to a weaker coordination of Pb-I bond. In addition, TOPO was reported to form a dipole layer at the surface of perovskite that points outwards and reduces the work function (in other words, shifts the vacuum level to lower energy) [22, 35].

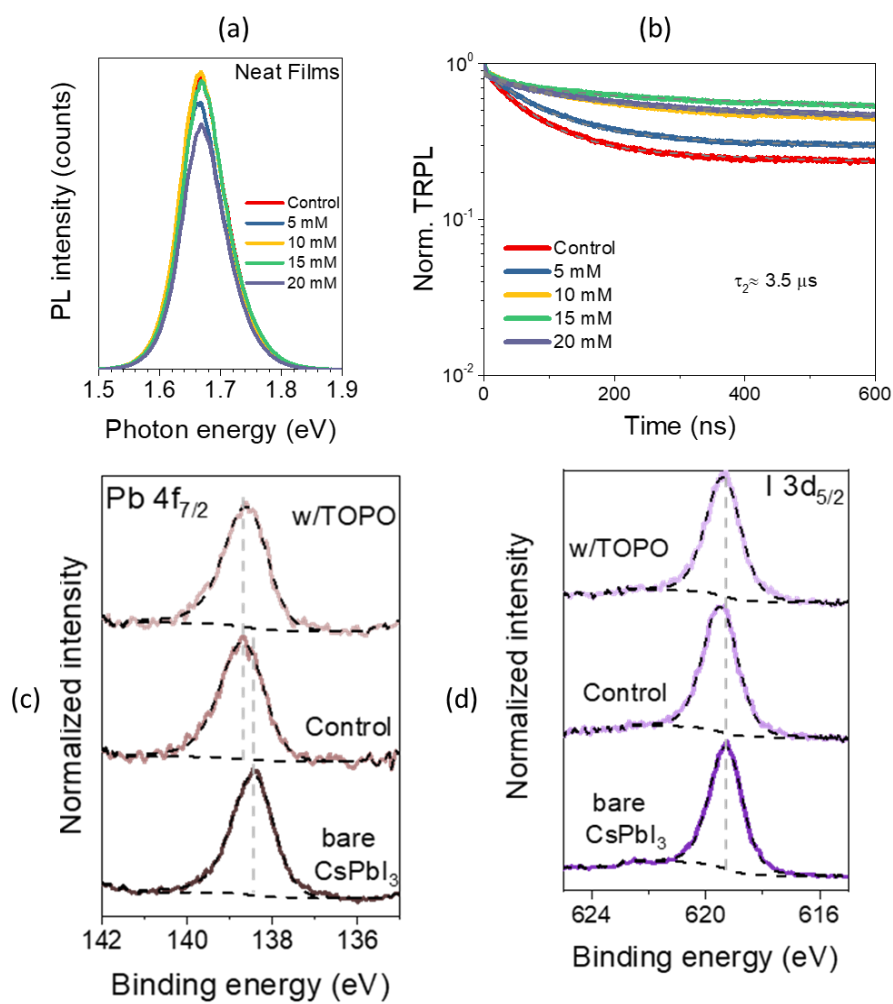

**Fig. 1 | Surface Passivation**. a) Steady-state and b) time-resolved photoluminescence of neat perovskite films deposited on quartz for control (red line), 5 mM (steal blue line), 10 mM (yellow line), 15 mM (green line), and 20 mM (lavender line) TOPO treated samples c) XPS of Pb $4f_{7/2}$ and d) I $3d_{5/2}$ core-shell spectra of bare CsPbI$_3$, control sample, and w/TOPO. Dashed lines in black are fitted curves and backgrounds with guidelines in grey.

The core level binding energy shift in Pb 4f and I 3d also implies the direct contact between TOPO and perovskite film in our work. It is reasonable as we do not expect to have a compact thin layer of OAI on top of perovskite film that will lead to the formation of an insulating layer. Rather OAI is expected to incorporate into the perovskite crystal structure and form a 2-dimensional structure on the surface.[28] We suggest that OAI treatment has passivated most of the under-coordinated Pb2+ at the perovskite surface, and TOPO donates electrons more likely to the interfacial iodine vacancies as they are found to be benign.[36] This would explain what we observed in PL analysis in **Fig. 1a** that, TOPO does not contribute noticeably to defects passivation. From the XPS in **Fig. S8**, we detected signals for P 2s and O 1s peaks originating from TOPO molecules for TOPO-treated samples.

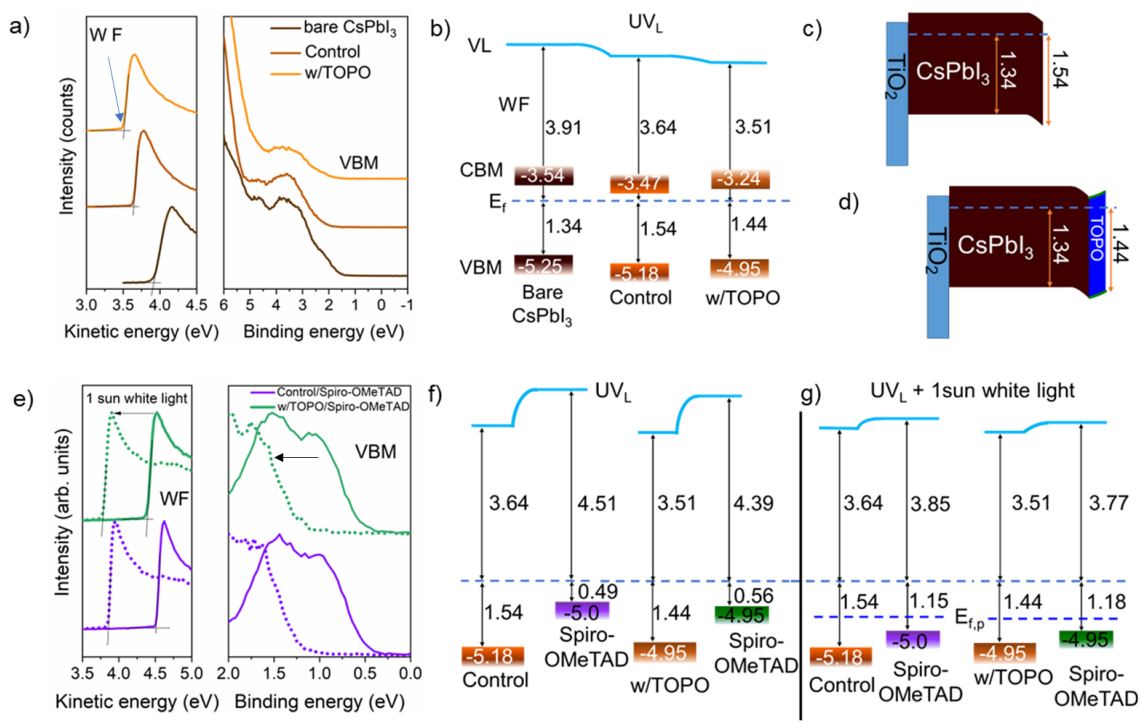

**Fig. 2 | Energy level alignment.** a) Ultraviolet photoelectron spectroscopy (UPS) spectra of bare CsPbI₃, control CsPbI₃, and with TOPO treatment: secondary electron cut-off (SECO, left panel) and valence spectra (right panel).Energetic level scheme of b)bare CsPbI₃, control CsPbI₃, and w/TOPO. Energy band diagram of c) control perovskite, d)with TOPO. e) UPS spectra of control/Spiro-OMeTAD, and w/TOPO/Spiro-OMeTAD: SECO (left panel) and valence spectra (right panel). Dashed lines are UPS spectra measured in the white light with 1 sun equivalent intensity. Energy band diagram of f) control/Spiro-OMeTAD and w/TOPO/Spiro-OMeTAD in the dark, and g) under 1 sun white light illumination. All samples were deposited on top of FTO/TiO₂. UV$_L$ refers to the low UV flux attenuated by the monochromator.

**Fig. 2a and e** (left panel) show the secondary electron cut-off spectra of all the samples, from which the work function (WF) can be obtained using this equation: $WF = hv - E_{cut-off}$, where $hv$ is the photoelectron energy of He (I) light, and $E_{cut-off}$ is the secondary electron cut-off. The right panel shows the valence band (VB) spectra of these samples on a linear intensity scale of the photoelectrons. It is noted that the VB onset of perovskites is, however, extrapolated on a logarithmic intensity scale (**Fig. S9**) to accurately infer the band edge position for perovskites. [37-38] We plotted the WF and VB onset of our samples in **Fig. 2b** with the energy band diagram illustrated in **Fig. 2c-d**. The conduction band minimum (CBM) was positioned given the optical bandgap that was obtained from the Tauc plot for control and TOPO-treated samples given in **Fig. S2b**. The values of WF, VBM (valence band maximum), and CBM are summarized in **Table S3**. We observed a decrease in WF in both OAI and TOPO-treated samples. This is further confirmed by Kelvin probe measurement (**Fig. S10, Table S4**).

Moreover, as discussed above, OAI treatment introduced a downward band bending of approximately 200 meV at the surface, resulting in a more *n*-type perovskite film which is in line with a recent report.[39] The addition of a TOPO layer, on the other hand, reversed the surface band bending into an upward one by around 100 meV. With the adjacent hole transport material (HTM), *i.e.* Spiro-OMeTAD in this work, we observed that the ground state interfacial energy levels exhibit dramatic changes in both samples, which is believed to be caused by charge carrier rearrangement at the interface of perovskite and HTM, as observed in our previous studies.[40,41] It is also noted that the abrupt increase in vacuum level at the interface is almost identical to both samples by around 870 meV. In order to approach the interfacial energy level alignment under device operating conditions, we further conducted the Ultraviolet photoelectron spectroscopy (UPS) measurement under the additional white light with light intensity equivalent to 1 sun (dashed line in **Fig. 2e**). The energy levels of the spiro-OMeTAD layer are found to exhibit dramatic downward shifts by 0.66 eV and 0.62 eV for the control and TOPO treated samples, respectively. Such shifts, as recently observed in perovskite/organic semiconductor interfaces, [40,41] are caused by charge carrier accumulation at the interface under illumination, which leads to a realignment of the energy levels at perovskite/Spiro-OMeTAD interface with spiro-OMeTAD HOMO shifting towards the perovskite VBM. It is noted that the measured energy levels from photoemission always refer to the Fermi level of the conductive FTO substrate. Given the unchanged perovskite energy levels upon white light illumination, the large energy offset between perovskite VBM and spiro-OMeTAD HOMO is then significantly reduced to 210 meV for control and to 260 meV for TOPO-treated samples. So far, we revealed two roles played by the TOPO layer. One is its function as a dipole pointing towards the perovskite film leading to a decrease in WF compared to the control sample, confirmed from both UPS (~130 meV) and Kelvin probe measurement (~ 210 meV). The influence of tuned WF on device stability was studied in one recent paper [39]. It was reported that a lower WF, *i.e.* a less negative vacuum level, reduces the halide migration activation energy and thus leads to more pronounced hysteresis and less device stability.[39] This is not what we observed in our work. We detected a decrease in WF for TOPO-treated samples. The other role of TOPO is the chemical bonding with surface ions in perovskite that leads to an upward surface band bending of 100 meV. This change in interfacial energy level alignment is in our great interest. In the following, we are going to discuss its effect on charge exaction and charge selectivity and, most importantly, on device stability.

**Interface charge selectivity**

Charge carrier selectivity at the interface of our samples was characterized using transient surface photovoltage spectroscopy (tr-SPV). The tr-SPV signals are directly proportional to the separated charges (SPV(t)~($n$(t)-$p$(t))×0.5L) at the buried interface of a device (L is the thickness of the light absorber layer, *i.e.* perovskite film in our study). Therefore, transient SPV provides important insights into the dynamics of charge extraction and recombination.[42,43] This technique was employed in our previous publication on the comparison of a range of self-assembled monolayers (SAM) comprising hole transport materials for the efficiency of hole extraction.[31] We measured transient SPV for several different samples, *i.e.* control perovskite films, TOPO-treated perovskite films, control perovskite films with spiro-OMeTAD, TOPO-treated perovskite films with spiro-OMeTAD (all deposited on glass/TiO$_2$). We also conducted the measurement for samples deposited on glass (**Fig. S11**). It shows that CsPbI$_3$ films deposited on glass have shallow and deep trap states (with activation energy up to 0.8 eV) while the TiO$_2$ substrate helps to remove these trap states and results in a clean and sharp SPV contour plots at a wide spectral range from 1.8 to 3.0 eV.

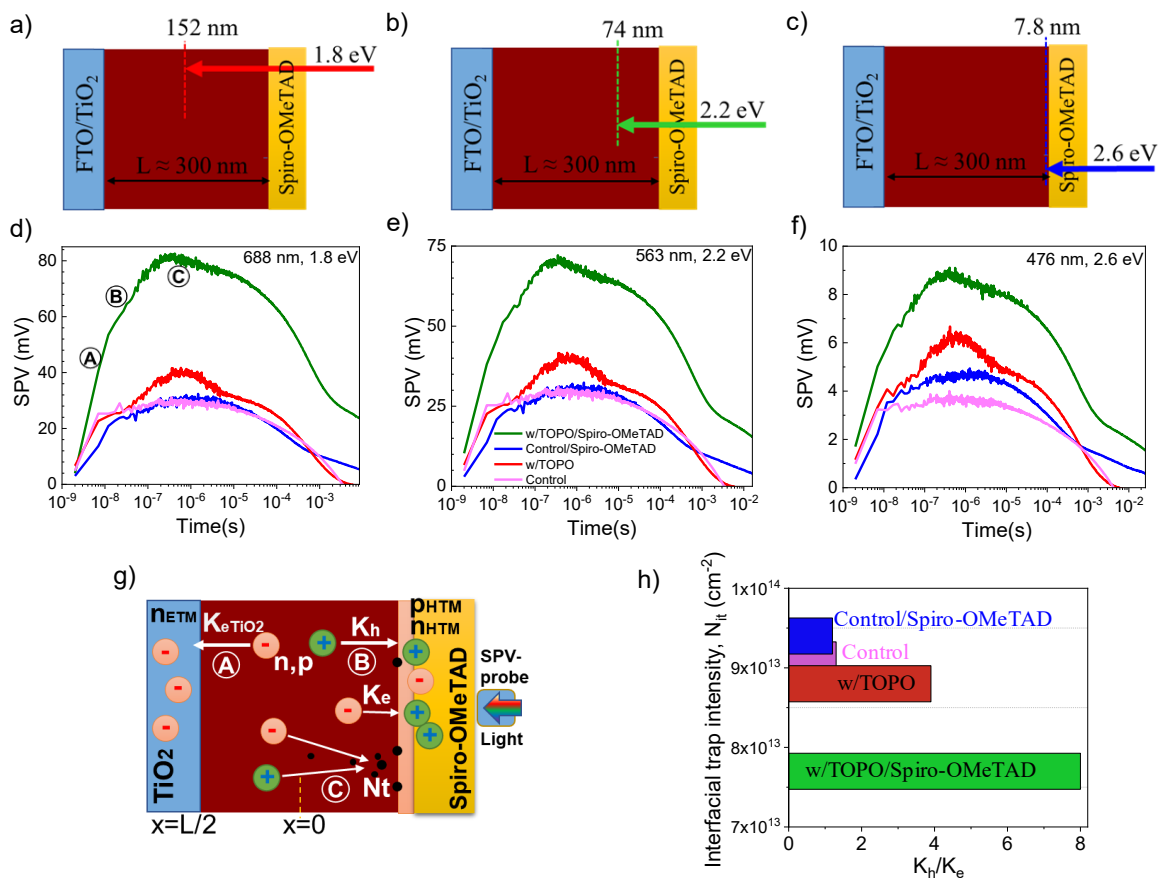

**Fig. 3 | Charge selectivity.** Scheme of light penetration depth into the samples for a) red light (688 nm, 1.8 eV), b) green light (563 nm, 2.2 eV), and c) blue light (476 nm, 2.6 eV). Transient SPV of control (pink line), w/TOPO (red line), control/Spiro-OMeTAD (blue line), w/TOPO/Spiro-OMeTAD (green line) measured at excitation source of d) 688 nm (1.8 eV), e) 563 nm (2.2 eV), and

f) 476 nm (2.6 eV) at fluence of 0.010 μJ/cm$^2$, 0.029 μJ/cm$^2$, and 0.040 μJ/cm$^2$, respectively, equivalent to 1sun light intensity. g) Fitted transient SPV data with h) charge extraction and recombination model describing carrier transport to ETL and HTM layers at two electrodes. The rate constant $K_{eTiO2}$ characterizes electron injection from perovskite to the TiO$_2$ layer (process A). The constants $K_e$ and $K_h$ correspond to electron and hole injection rates from the perovskite to the HTM side (process B). Defect concentration $N_t$ is responsible for SRH non-radiative recombination (process C). The extracted electron and hole concentrations $n_{etm}$ and $p_{htm}$ induce the simulated tr-SPV signal as shown in **Fig. S15** by black curves (more details in Eq. S7). i) Ratio of $K_h$ to $K_e$ and its correlation to interfacial trap density, extracted from the fitting.

**Fig. 3a-c** exhibit the penetration depth of three different photon energies used in tr-SPV characterization. The penetration depth is calculated from the absorption coefficient of perovskite films (**Fig. S2d**) as the top HTM has an absorption onset of approximately 2.92 eV [44]. It shows that at red light (1.8 eV) excitation, it can penetrate close to the middle of the film. At the green light (2.2 eV) excitation, it penetrates approximately 74 nm deep into the film. At the blue light (2.6 eV), it penetrates within 10 nm next to the interface. **Fig. 3d-f** presents the tr-SPV results of samples measured under these three photon energies at fluences of 0.010 μJ/cm$^2$, 0.029 μJ/cm$^2$, and 0.040 μJ/cm$^2$, respectively, corresponding to the carrier concentration of 1.4×10$^{15}$ cm$^{-3}$ close to 1 sun operation conditions. Such a measurement was also conducted at 0.1 sun and 10 suns equivalent (**Fig. S12**), as discussed in SI. Thus, the larger photon energies (i.e. 2.2 eV and 2.6 eV) generated charge carriers closer to the HTM. This includes free carrier nonuniformities in the perovskite film, demands longer charge diffusion distance, and results in lower SPV amplitude in the same samples. However, we observed the same trend for the measurement conducted at three photon energies and three light intensities TOPO-treated samples have higher SPV amplitude than the control samples.

In general, the SPV signal for efficient HTM (or ETM) appears as a fast exponential rise with a large amplitude. The recombination and non-efficient charge selectivity tend to slow down the rise and decrease the SPV amplitude. We observe a significant boost of hole extraction in the TOPO-treated sample in the presence of adjacent HTM, approximately two times higher in the SPV amplitude and much faster signal rise than that of control samples in all studied laser photon energies. In contrast, for control samples, with and without spiro-OMeTAD HTM, the SPV amplitude is almost the same. The rise in SPV amplitude in the control sample with HTM starts to be noticed under blue light illumination. In other words, for control samples, holes can be properly extracted only in the vicinity of HTM. Furthermore, TOPO-treated samples even without HTM showed a higher SPV amplitude than the control and with HTM. This behaviour of TOPO-treated devices implies the positive effects of TOPO treatment on free electron extraction, hole extraction, and charge recombination.

To explain in detail, charge extraction dynamics, we developed a 1D simulation of charge. We will mainly focus here on the processes that occurred during the 1 ns-10 μs timescale (four orders of magnitudes) associated with carrier extraction and recombination. The initial charges are generated in the perovskite layer by the light and can be extracted to ETM and HTM with extraction rate constants $K_{eETM}$ and $K_h$, respectively, **Fig. 3g**. To describe charge selectivity properties at HTM we also introduced electron injection rate constant $K_e$ responsible for electron injection to HTM (for example, as a result of poor selectivity). The non-radiative and radiative charge recombinations are characterized by defect concentration $N_T$ and band-to-band recombination $C_b$. To simplify the fitting procedure, we adopted part of the constants from the literature [45,46] (see **Table S5**). More details on the model can be found in SI. The results of the fit and main fitting parameters are summarized in **Fig. S13** (black curves) and **Table S6**.

According to the model results, the fast initial rise (region (A) in **Fig. S13**) is formed due to rapid electron extraction to $TiO_2$ with the rate constant ($K_{eETM}$) of $1.8 \times 10^7$ s$^{-1}$. The electron extraction rate constant only slightly decreased to $1.3 \times 10^7$ s$^{-1}$ in other devices, signaling that other processes are responsible for dramatic carried dynamics changes in time regions (B) and (C). The difference in $K_{eETM}$ can originate from a slightly faster electron extraction at the interface of $TiO_2$ and $CsPbI_3$ due to enhanced energy offset at the CBM level as characterized by the UPS measurement.[44]

We further found that TOPO treatment significantly boosts the free holes extraction rate ($K_h$) and selectivity properties of HTM. The value of $K_h$ increased nearly twice in TOPO-treated samples with spiro-OMeTAD ($8 \times 10^6$ s$^{-1}$) compared with the control sample ($4.4 \times 10^6$ s$^{-1}$). The improved hole extraction near the TOPO surface is also observed in TOPO-treated perovskite films ($3.9 \times 10^6$ s$^{-1}$) in comparison with the control sample ($1.5 \times 10^6$ s$^{-1}$). In addition to improved hole extraction, TOPO boosts significantly the selectivity properties of HTL, which can be characterized by $K_h/K_e$ ratio where $K_e$ is electron injection in HTM (**Table S6**). **Fig. 3h** shows that TOPO treatment increased the $K_h/K_e$ ratio from the value of 1.3 in control to the value of 3.9 for perovskite films. Further, this value increased up to 8 in the presence of spiro-OMeTAD, meanwhile, control samples without TOPO treatment had a rather low value of 1.2. Finally, we observe slight passivation of non-radiative recombination in TOPO-treated samples with spiro-OMeTAD (interfacial trap density of $7.7 \times 10^{13}$ cm$^{-3}$) in comparison with control samples with spiro-OMeTAD ($9.4 \times 10^{13}$ cm$^{-3}$). This is in good alignment with what we observed in the PL measurement as presented in **Fig. 1a**.

The overall picture of the charge dynamics suggested by tr-SPV with simulation demonstrates a significant role of TOPO in the improvement of hole extraction. The TOPO layer stimulates free hole extraction in HTM and repels electrons from the TOPO surface. The selectivity of neat HTM can result in noticeable suppression of charge recombination near the HTM surface and in HTM itself. TOPO also provides chemical passivation of traps (18% decrease in trap concentration). We believe that the suppressed recombination is mainly contributed by the enhanced hole selectivity due to the formation of upward surface band bending at the interface between CsPbI$_3$ and spiro-OMeTAD. The enhanced hole selectivity originates from better energetic alignment caused by upward surface band bending and dipole activity which contributes to recombination suppression as well.

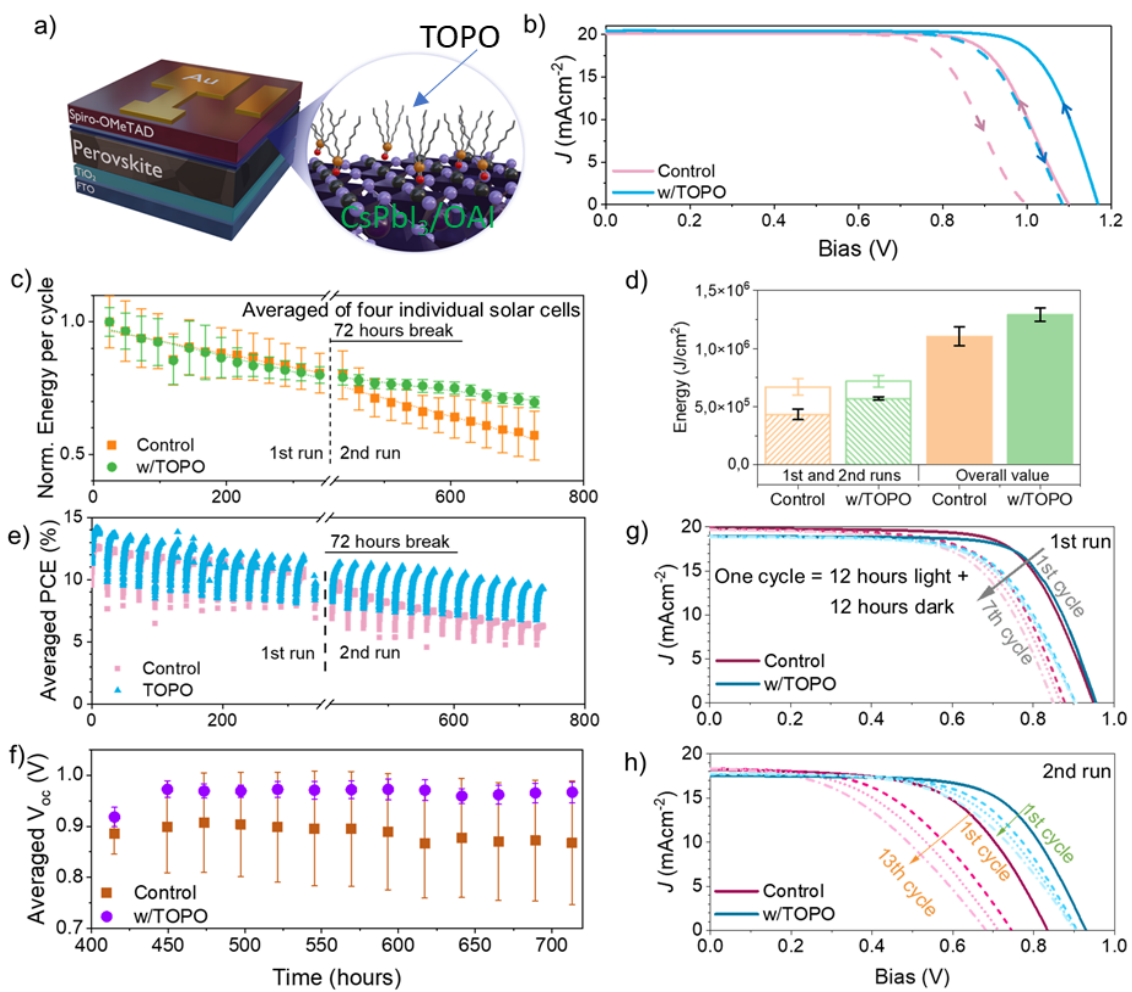

**Fig. 4 | *J-V* measurements and Device stability.** a) Scheme of the device structure with the highlighted part for the TOPO treatment on the perovskite surface. b) The champion *J-V* curve (100 mV/cm$^2$, AM1.5G, scan rate: 200 mV/s, reverse scan and forward scan indicated as the solid and dash line, respectively, and further indicated in arrows), c) Normalized energy output per cycle, with each cycle composed of 12 hours light-on and 12 hours of light-off, d) energy output in the first and second runs and the overall value, e) averaged efficiency, f) evolution of averaged $V_{oc}$ as a function

of ageing hours in the second run. f) are averaged from four individual solar cells. *J-V* curves evolution as a function of illumination cycles in g) the first run and h) the second sun for the control and TOPO-treated samples. The dashed lines in g) and h) refer to the number of illumination cycles, with the last cycle at 13 for both samples.

**Fig. 4a** shows the scheme of the device structure with a compact $TiO_2$ layer as the electron transport material (ETM) and spiro-OMeTAD as the hole transport material (HTM). The TOPO molecule is represented on top of perovskite film with the alky chains visible. All the samples, with and without the TOPO, are treated with OAI to passivate the defects of the perovskite, as reported previously.[24] The champion device *J-V* curves in **Fig. 4b** show a clear improvement of $V_{oc}$ with the TOPO treatment. The improved hole extraction is found to significantly improve the $V_{oc}$ and the *FF* of the cells as presented in the box chart in **Fig. S15**. Moreover, we achieved a $V_{oc}$ of over 1.2 V under the TOPO treatment, while control samples showed a $V_{oc}$ strictly below 1.2 V, similar to what was reported in the reference.[33,47] *J-V* curves with $V_{oc}$ of over 1.2 V are given in **Fig. S16**. We observed robust $V_{oc}$ and *FF* of TOPO-treated solar cells in the long-term stability test as well, as presented in **Fig. 4f** and **Fig. S17-I d**.

We focused on the stability of $CsPbI_3$, which is more critical than the organic-inorganic perovskite compositions.[29,48] The statistical analysis of the device performance parameters is reported in the SI, and it is consistent with the performance of the champion devices. Device stability was measured in a custom-built MPPT (maximum power point tracking) ageing system[49] according to ISOS-LC-11[6], by alternating 12 hours of illumination under one sun and 12 hours of dark for 27 cycles separated in two runs with a break of 72 hours. We conducted the stability test in nitrogen and at ambient temperature of 25 °C including the 72 hours break in the dark phase for all measured samples. More details of the long-term stability test can be found in the SI. The UV component of the solar spectrum was filtered to prevent degradation caused by $TiO_2$ oxygen desorption.[50,51] **Fig. 4c** shows the energy produced per cycle, normalized to the first cycle. Each point is an average of the energy produced by four independent devices (the individual data of each sample and the applied method to filter data, as well as a full statistic evaluation, are given in SI and **Fig. S17a-b**). In the first run (*i.e.* 14 cycles in 342 hours), we observe a similar degradation trend for both samples, while a much slower degradation process is observed in TOPO-treated samples in the second run (*i.e.* 15th-27th cycles in 322 hours). The overall energy generated by the solar cells is averaged and plotted in **Fig. 4d**. Overall, TOPO-treated solar cells can generate 16.8% more energy than the control, with the surplus energy mainly generated after the 342 hours (33.3% over the control). It should be noted that control samples have a larger deviation than TOPO treated ones. We took one representative sample from each group and plotted PCE normalized to the initial value in **Fig. 4e**. The evolution of PCE follows the same trend as the produced energy in **Fig. 4c**.

**Fig. 4f** shows that a significantly more stable $V_{oc}$ is observed in TOPO-treated samples compared to the control since the 414[th] hour of the stability measurement. **Fig. 4g** and **4h** present the *J-V* curve evolution of control and TOPO-treated samples during the first and the second runs, respectively. We can see that both samples exhibit good stability in the initial cycles. Yet the superior stability in TOPO-treated samples becomes more pronounced from the 15[th] cycle. In the overall 14 cycles in the second run, the control sample degrades significantly with a drop mainly in $V_{oc}$. In contrast, TOPO-treated samples exhibit only a slight decrease in $V_{oc}$. The stability of the control in the first 342 hours is consistent with the literature that used OAI to passivate the defects of the perovskite. Nevertheless, on a longer timescale, the perovskite passivation solo is not sufficient to guarantee stability. The reason behind the improved stability of the TOPO sample will be investigated in the following paragraph.

So far, we have demonstrated that better charge selectivity leads to better device stability. The reasoning behind this can be understood in the following. We noted a much more severe $V_{oc}$ loss **(Fig. 4h)** and FF loss **(Fig. S17-I-d)** in the *J-V* curves of control samples, with respect to the TOPO samples, recorded from the 1[st] to the 13[th] cycle during the aging test. In particular, the *FF* loss is mainly coming from the increase in serial resistance as the slope of the *J-V* curve at the bias around the $V_{oc}$ becomes less and less steep over the cycled illumination. The increase in serial resistance implies less conductivity in contact layers, such as the ETM, HTM, or metal electrode. We suggest that it is caused by iodide migration across the interface to the HTM layer during illumination and under applied bias.[52] It was reported that the chemical reaction happens between spiro-OMeTAD[+] and migrating I[-], which progressively reduces the conductivity in HTM. [52,53] With the TOPO treatment, the resulting upward band bending due to the chemical bonding increases the energy barriers for electrons and negatively charged mobile ions, such as iodide, to move across the interface. This prevents film conductivity drops of HTM and thus holds the initial *FF* robustly in TOPO samples. Besides, the iodide diffusion into spiro-OMeTAD will shift up the HOMO levels towards the vacuum level because the radical concentration of spiro-OMeTAD[+] will be reduced due to the reaction with iodide and that leads to less *p*-doping in spiro-OMeTAD. Then the interfacial energetic level re-alignment will create more non-radiative recombination, which leads to severe $V_{oc}$ loss.[54]

Additionally, by chemical bonding with iodine vacancies at the interface, TOPO helps to make iodine vacancies less mobile. Iodine vacancies at the interface are found to be benign but illumination can promote the diffusion of iodine vacancies to the bulk that makes them detrimental for creating new non-radiative recombination centers. [55] This causes severe loss in $V_{oc}$ in control samples but is effectively suppressed in TOPO samples. Thirdly, the energetically upward band bending enhances the charge selectivity, *i.e.* better separation between electrons and holes at the interface, which leads to less charge accumulation near the interface and thus less interfacial recombination as supported by tr-SPV data in **Fig. 3**. In return, it contributes to better stability in devices.[56] The synergy of the above effects of TOPO provides better device stability and performance in TOPO-treated samples.

**Conclusions:**

Herein, we developed a strategy based on surface dipole-induced band bending to mitigate the charge loss at the perovskite/HTL interface. We have demonstrated that the TOPO molecule causes surface upward band bending and resolves unfavorable downward bending induced by OAI passivation of the $CsPbI_3$ film. The TOPO dipole molecule shifts the interfacial energy levels and reduces the energy offset between perovskite VBM levels and spiro-OMeTAD HOMO levels shown by UPS measurements. Further, this energy level alignment induces a boost in charge transfer at the interface as revealed by the trSPV study. The competing effect of passivation and charge extraction is resolved by charge transport simulation proving that TOPO plays a minor role in trap passivation and mainly significantly boosts the hole extraction rate. By applying the TOPO dipole at the HTL interface in an all-inorganic perovskite solar cell, we have achieved a $V_{oc}$ of over 1.2V with improved stability and 19% efficiency, and 16% more energy production for the TOPO-treated samples. This work reveals that besides defects passivation interfacial energy level alignment and charge selectivity play a critical role in device stability and efficiency.

**Conflict of Interest**

The authors declare no conflict of interest.

**Acknowledgment**

Z.I. acknowledges the Deutscher Akademischer Austauschdienst (DAAD) scholarship for the financial support for his Ph.D. study at HZB. He also acknowledges the generous help from Dr. Marion Flakten for the XRD measurement, and Mrs. Carola Kilmm for the SEM characterization. T.W.G. appreciates the help from Dr. Fluorine Ruska in the UV-Vis measurement, and Dr. Nikolai Severin for AFM characterization. F. Z. and N. K. acknowledge the funding from Deutsche Forschungsgemeinschaft (DFG, German Research Foundation, Project numbers: 182087777-SFB951 and 423749265-SPP2196 "SURPRISE"). This project has received funding from the European Union's Framework Programme for Research and Innovation HORIZON EUROPE (2021-2027) under the Marie Skłodowska -Curie Action Postdoctoral Fellowships (European Fellowship) 101061809 HyPerGreen. We gratefully acknowledge Dr. Thomas Dittrich for providing the HZB SPV lab facilities We acknowledge HyPerCells (a joint graduate school of the University of Potsdam and the Helmholtz-Zentrum Berlin). M.S. acknowledges the Deutsche Forschungsgemeinschaft (DFG, German Research Foundation) - project numbers 423749265 and 424709669 - SPP 2196 (SURPRISE and HIPSTER) for funding and further financial support from the Federal Ministry for Economic Affairs and Energy within the framework of the 7th Energy Research Programme (P3T-HOPE, 03EE1017C).

**Authors contribution**

Z.I., Q.W., and A.A. perceived the idea. Q.W. and Z.I. designed the experiment. Z.I. made all the samples and solar cells involved in this work. F.Z. conducted XPS and UPS measurements and contributed to data analysis together with N.K. A.M. conducted tr-SPV and Kelvin probe measurements, and theoretical simulation of charge transport, and contributed to data analysis and manuscript proofreading and revision. E.G.P. conducted PL and trPL spectroscopy measurements and contributed to data analysis together with M.S. and D.N. T.W.G. conducted UV-Vis spectroscopy and AFM measurements for this project. H.K. measured the long-term stability of solar cells and contributed to data analysis together with Q.W. and A.A. L.C. designed the device stack scheme, and was involved in the discussion on dipole interaction with perovskite film in this project. G.S did some device measurements and discussed with M.P. and A.B.M.G . Z.I. and Q.W. worked on the manuscript together. The final draft of the manuscript has been discussed with all co-authors.